\documentstyle[12pt]{article}
\begin{document}
\begin{titlepage}
\begin{flushright}
NTUA--03/02 \\ hep-ph/0203095
\end{flushright}
\vspace{1cm}

\begin{centering}
\vspace{.4in} {\Large {\bf Soft Supersymmetry Breaking from\\
\vspace{.3cm} Coset Space Dimensional Reduction.}}\\
\vspace{1.5cm}

{\bf P.~Manousselis}$^{a}$ and {\bf G.~Zoupanos}$^{b}$\\
\vspace{.2in} Physics Department, National Technical University,
\\ Zografou
Campus, 157 80 Athens, Greece.\\

\vspace{1.0in}

{\bf Abstract}\\

\vspace{.1in} The Coset Space Dimensional Reduction scheme is
briefly reviewed. Then a ten-dimensional supersymmetric $E_8$
gauge theory is reduced over symmetric and non-symmetric
six-dimensional coset spaces. In general a four-dimensional
non-supersymmetric gauge theory is obtained in case the used
coset space is symmetric, while a softly broken supersymmetric
gauge theory is obtained if the used coset space is
non-symmetric. In the process of exhibiting the above properties
we also present two attractive models, worth exploiting further,
which lead to interesting GUTs with three families in four
dimensions.
\end{centering}
\vspace{4.7cm}

\begin{flushleft}
$^{a}$e-mail address: pman@central.ntua.gr. Supported by
$\Gamma\Gamma$ET  grand 97E$\Lambda$/71.
\\ $^{b}$e-mail address:
George.Zoupanos@cern.ch. Partially supported by EU under the RTN
contract HPRN-CT-2000-00148 the A.v.Humboldt Foundation and the
Greek-German Bilateral Programme IKYDA-2001.
\end{flushleft}
\end{titlepage}

\section{Introduction}
In the recent years the theoretical efforts to establish a deeper
understanding of Nature have managed to achieve a number of
successes in developing frameworks that aim to describe the
fundamental theory at the Planck scale. On the other hand a real
breakthrough in a deeper knowledge of Nature should include an
understanding of the, at present, free parameters of the Standard
Model (SM) in terms of a few fundamental ones. Clearly the
apparent success of the SM is spoiled by the presence of a
plethora of free parameters mostly related to the ad-hoc
introduction of the Higgs and Yukawa sectors in the theory. It is
worth recalling that the Coset Space Dimensional Reduction (CSDR)
\cite{Witten,Manton,Kuby,Review} was suggesting from the beginning
that a unification of the gauge and Higgs sectors can be achieved
in higher dimensions. The four-dimensional gauge and Higgs fields
are simply the surviving components of the gauge fields of a pure
gauge theory defined in higher dimensions. In the next step of
development of the CSDR scheme, fermions were introduced
\cite{Slansky,Palla} and then the four-dimensional Yukawa and
gauge interactions of fermions found also a unified description in
the gauge interactions of the higher-dimensional theory. The last
step in this unified description in high dimensions is to relate
the gauge and fermion fields that have been introduced. A simple
way to achieve that is to demand that the higher-dimensional gauge
theory is\ ${\cal N}= 1$ supersymmetric which requires that the
gauge and fermion fields are members of the same supermultiplet.

In the spirit described above a very welcome additional input is
that string theory suggests furthermore the dimension and the
gauge group of the higher-dimensional supersymmetric theory
\cite{Theisen}. Further support to this unified description comes
from the fact that the reduction of the theory over coset
\cite{Review} and CY spaces \cite{Theisen} provides the
four-dimensional theory with scalars belonging in the fundamental
representation of the gauge group as are introduced in the SM. In
addition the fact that the SM is a chiral theory leads us to
consider $D$-dimensional supersymmetric gauge theories with
$D=4n+2$ \cite{Chapline,Review}, which include the ten dimensions
suggested by the heterotic string theory \cite{Theisen}.

Concerning supersymmetry, the nature of the four-dimensional
theory depends on the corresponding nature of the compact space
used to reduce the higher-dimensional theory. Specifically the
reduction over CY spaces leads to supersymmetric theories
\cite{Theisen} in four dimensions, the reduction over symmetric
coset spaces leads to non-supersymmetric theories, while a
reduction over non-symmetric ones leads to softly broken
supersymmetric theories \cite{Pman1,Pman2,Megalo}. Concerning the
latter as candidate four-dimensional theories that describe the
nature, in addition to the usual arguments related to the
hierarchy problem, we should remind a further evidence established
in their favor the last years. It was found that the search for
renormalization group invariant (RGI) relations among parameters
of softly broken supersymmetric GUTs considered as a unification
scheme at the quantum level, could lead to successful predictions
in low energies. More specifically the search for RGI relations
was concerning the parameters of softly broken GUTs beyond the
unification point and could lead even to all-loop finiteness
\cite{Mondragon1,Mondragon2}. On the other hand in the low
energies lead to successful predictions not only for the gauge
couplings but also for the top quark mass, among others, and to
interesting testable predictions for the Higgs mass
\cite{Kobayashi}.

The review is organized as follows. In section~\ref{sec:CSDR} we
describe the CSDR. In section~\ref{sec:symm} we describe two
representative examples of CSDR of a ten-dimensional
supersymmetric $E_{8}$ gauge theory over symmetric coset spaces.
In section~\ref{sec:non-symm} we present the CSDR of the same
$E_{8}$ gauge theory over the three existing non-symmetric coset
spaces. Finally section~\ref{sec:concl} contains our conclusions.
\section{The Coset Space Dimensional Reduction (CSDR)
scheme}\label{sec:CSDR} Let us proceed by recalling the
fundamental ideas of the dimensional reduction procedure.
Dimensional reduction is the construction of a lower dimensional
Lagrangian starting from higher dimensions. In our construction we
will  use the symmetries of the extra dimensions. The CSDR is a
dimensional reduction scheme where the extra dimensions form a
coset space $S/R$ and the symmetry used is the group $S$ of
isometries of $S/R$.

Given this form of the extra coordinates one can construct a lower
dimensional Lagrangian by demanding the fields to be symmetric.
This means that one requires the fields to be form invariant under
the action of the group $S$ on extra coordinates,
\begin{equation}\label{eq:frminv}
\delta_{\xi}\Phi^{i}= L_{\xi}\Phi^{i} =0,
\end{equation}
where $L_{\xi}$ is the Lie derivative with respect to the Killing
vectors $\xi$ of the extra dimensional metric. However this
condition is too strong when the higher-dimensional Lagrangian
possesses a symmetry. Then a generalized form of the symmetry
condition~(\ref{eq:frminv}) can be
\begin{equation}
\delta_{\xi}\Phi^{i}=L_{\xi}\Phi^{i} = U_{\xi}\Phi^{i},
\end{equation}
where $U_{\xi}$  is the symmetry of the Lagrangian.

When we apply CSDR in a gauge theory the $U_{\xi}$ is a gauge
transformation. Then the original  Lagrangian becomes independent
of the extra coordinates because of gauge invariance.

To be specific we consider a Yang-Mills-Dirac Lagrangian with
gauge group $G$ in $D$ dimensions
\begin{eqnarray}
S&=&\int d^{D}x\sqrt{-g}\Bigl[-\frac{1}{4}
Tr\left(F_{MN}F_{K\Lambda}\right)g^{MK}g^{N\Lambda}\nonumber\\
&&+\frac{i}{2}\overline{\psi}\Gamma^{M}D_{M}\psi\Bigr] ,
\end{eqnarray}
where capital indices $M,N$ run from $0 \ldots D-1$. The field
strength is
\begin{equation}
F_{MN}=\partial_{M}A_{N}-\partial_{N}A_{M}-\left[A_{M},A_{N}\right],
\end{equation}
with $A_{M}$ the gauge field. $\Gamma^{M}$ are the $D$-dimensional
gamma matrices satisfying
\begin{equation}
\{ \Gamma_{M},\Gamma_{N}\}=2g_{MN},
\end{equation}
$\psi$ is an anticommuting spinor in $D$ dimensions and
$\overline{\psi}$ is the Dirac adjoint. $D_{M}$ is the covariant
derivative,
\begin{equation}
D_{M}= \partial_{M}-\theta_{M}-A_{M},
\end{equation}
with
\begin{equation}
\theta_{M}=\frac{1}{2}\theta_{MN\Lambda}\Sigma^{N\Lambda},
\end{equation}
the spin connection of $M^{D}$.

Then the original spacetime $(M^{D},g^{MN})$, is assumed to be
compactified to $M^{4}~\times~S/R$, with $S/R$ a coset space. The
original  coordinates $z^{M}$ become coordinates of $M^{4} \times
S/R$, $z^{M}= (x^{m},y^{\alpha})$,where $\alpha$ is a curved index
of the coset, and with $a$ we will denote a flat tangent space
index. The metric is
\begin{equation}
g^{MN}= \left[\begin{array}{cc}\eta^{\mu\nu}&0\\0&-g^{\alpha
\beta}\end{array} \right],
\end{equation}
where $\eta^{\mu\nu}~=~diag(1,-1,-1,-1)$ is the Minkowski
spacetime metric and $g^{\alpha \beta}$ is the coset space metric.

To develop the geometry of coset spaces we divide the generators
of $S$, $ Q_{A}$ in two sets, the generators of $R$, $Q_{i}$
$(i=1, \ldots,dimR)$ and the generators of $S/R$, $ Q_{a}$
$(a=dimR+1 \ldots,dimS)$. Then the commutation relations for the
$S$ generators become
\begin{eqnarray}
\left[ Q_{i},Q_{j} \right] &=& f_{ij}^k Q_{k},\nonumber \\ \left[
Q_{i},Q_{a} \right]&=& f_{ia}^{b}Q_{b},\nonumber\\ \left[
Q_{a},Q_{b} \right]&=& f_{ab}^{i}Q_{i}+f_{ab}^{c}Q_{c} .
\end{eqnarray}
The coset space $S/R$ is a symmetric one when $f_{ab}^{c}=0$.

The coordinates $y$ define an element of $S$, $L(y)$, which is a
coset representative. Then the Maurer-Cartan form with values in
the Lie algebra of $S$ is defined by
\begin{equation}
V(y)=L^{-1}(y)dL(y) = e^{A}_{\alpha}Q_{A}dy^{\alpha}
\end{equation}
and obeys the Maurer-Cartan equation,
\begin{equation}\label{eq:MC}
dV+V \wedge V=0.
\end{equation}
From the relation (\ref{eq:MC}), and using standard techniques of
differential geometry we can develop the geometry of coset spaces,
and compute vielbeins, connections, curvature and torsion. For
instance the vielbein and the $R$-connection can be computed at
the origin $y=0$, and the results are, $ e^{a}_{\alpha} =
\delta^{a}_{\alpha}$ and $e^{i}_{\alpha} = 0$.

Next we require that the fields are invariant up to a gauge
transformation when $S$ acts on $S/R$. Let $\xi_{A}^{\alpha}$, $A
=1,\ldots,dimS$, be the Killing vectors which generate the
isometries $S$ of $S/R$ and denote by $W_{A}$ the compensating
gauge transformation associated with $\xi_{A}$. Define, next, the
infinitesimal coordinate transformation
\begin{equation}
\delta_{A} \equiv L_{\xi_{A}}.
\end{equation}
Then the condition that a coordinate transformation is compensated
by a gauge transformation takes the following specific form when
it is applied to the higher-dimensional vector and spinor
\begin{eqnarray} \label{cct1}
\delta_{A}A_{M}&=&\xi_{A}^{\beta}\partial_{\beta}A_{M}+\partial_{M}
\xi_{A}^{\beta}A_{\beta}\nonumber\\&=&\partial_{M}W_{A}-[W_{A},A_{M}],
\end{eqnarray}
\begin{eqnarray}\label{cct2}
\delta_{A}\psi&=&\xi_{A}^{\alpha}\partial_{\alpha}\psi-\frac{1}{2}G_{Abc}\Sigma^{bc}\psi
\nonumber\\&=& D(W_{A})\psi.
\end{eqnarray}
Inspecting eq.~(\ref{cct1}) we recognize the familiar form of the
Lie derivative of a vector and the gauge transformation
appropriate for the adjoint representation. The same is true for
the spinor in eq.~(\ref{cct2}). There $D(W_{A})$ represents a
gauge transformation in the representation to which the spinor
belongs. The quantities $W_{A}$ may depend on internal coordinates
$y$. The conditions (\ref{cct1}), (\ref{cct2}) should be covariant
when $A_{M}(x,y)$ and $\psi(x,y)$ transform under a gauge
transformation, therefore $W$'s transform under a gauge
transformation as
\begin{equation}\label{Wtr}
\widetilde{W}_{A}=gW_{A}g^{-1}+(\delta_{A}g)g^{-1}.
\end{equation}
Eqs~(\ref{cct1}),~(\ref{cct2}) and (\ref{Wtr}) provide us with the
gauge freedom to do all calculations at $y=0$ and $W_{a}=0$. Note
also that the variations $\delta_{A}$ satisfy
$[\delta_{A},\delta_{B}]=f_{AB}^{\\C}\delta_{C}$ leading to a
further consistency condition for the $W$'s
\begin{equation}
\xi_{A}^{\alpha}\partial_{\alpha}W_{B}-\xi_{B}^{\alpha}\partial_{\alpha}
W_{A}-\left[W_{A},W_{B}\right]=f_{AB}^{\ \ C}W_{C}.
\end{equation}

The above conditions imply certain constraints that the
$D$-dimensional fields should obey.

The solution of the constraints provide
\begin{itemize}
\item The remaining gauge invariance in four dimensions,
\item The four-dimensional spectrum,
\item The scalar potential of the four-dimensional theory.
\end{itemize}

More specifically to find the four-dimensional gauge group we
embed $R$ in $G$
\begin{eqnarray}
G &\supset& R_{G} \times H,\nonumber \\
H&=&C_{G}(R_{G}).
\end{eqnarray}
Then $H$, the centralizer of the image of $R$ in $G$, is the
four-dimensional gauge group.

The scalar fields that are obtained from the higher components of
the vector field are obtained as follows. First we embed  $R$ in
$S$  and decompose the adjoint of $S$ under $R$
\begin{eqnarray} \label{dec1}
S &\supset& R \nonumber \\ adjS &=& adjR+\sum s_{i}.
\end{eqnarray}
Then we embed $R$ in $G$ and decompose the adjoint of $G$ under $R
\times H $,
\begin{eqnarray} \label{dec2}
G &\supset& R_{G} \times H \nonumber \\
 adjG &=&(adjR,1)+(1,adjH)\nonumber\\&&+\sum(r_{i},h_{i}).
\end{eqnarray}
The rule is that when $$s_{i}=r_{i},$$ i.e. when we have two
identical representation of $R$ in the decompositions
(\ref{dec1}), (\ref{dec2}) there is an $h_{i}$ multiplet of scalar
fields that survives in four dimensions.

To find the representation of $H$ under which the four-dimensional
fermions transform, we have to decompose the representation $F$ of
$G$ to which the fermions belong, under $R_{G} \times H$, i.e.
\begin{equation}\label{eq:F}
F= \sum (t_{i},h_{i}),
\end{equation}
and the spinor of $SO(d)$ under $R$
\begin{equation}\label{eq:d}
\sigma_{d} = \sum \sigma_{j}.
\end{equation}
Then for each pair $t_{i}$ and $\sigma_{i}$, where $t_{i}$ and
$\sigma_{i}$ are identical irreducible representations, there is
an $h_{i}$ multiplet of spinor fields in the four-dimensional
theory.

In order to obtain chiral fermions in four-dimensions we have to
impose the Weyl condition in $D$ dimensions. In $D = 4n+2$
dimensions, which is the case of interest, the decomposition of
the left handed, say spinor under $SU(2) \times SU(2) \times
SO(d)$ is
\begin{equation}
\sigma _{D} = (2,1,\sigma_{d}) + (1,2,\overline{\sigma}_{d}).
\end{equation}
So we have in this case the decompositions
\begin{equation}
\sigma_{d} = \sum \sigma_{k},~\overline{\sigma}_{d}= \sum
\overline{\sigma}_{k}.
\end{equation}
Let us start from a vector-like representation $F$ for the
fermions. In this case each term $(t_{i},h_{i})$ in
eq.(\ref{eq:F}) will be either self-conjugate or it will have a
partner $( \overline{t}_{i},\overline{h}_{i} )$. According to the
rule described in eqs.~(\ref{eq:F}), (\ref{eq:d}) and considering
$\sigma_{d}$ we will have in four dimensions left-handed fermions
transforming as $ f_{L} = \sum h^{L}_{k}$. Since $\sigma_{d}$ is
non self-conjugate, $f_{L}$ is non self-conjugate too. Similarly
from $\overline{\sigma}_{d}$ we will obtain the right handed
representation $ f_{R}= \sum \overline{h}^{R}_{k}$ but as $F$ is
vector-like, $\overline{h}^{R}_{k}\sim h^{L}_{k}$. Therefore there
will appear two sets of Weyl fermions with the same quantum
numbers under $H$. This is already a chiral theory, but still one
can go further and try to impose the Majorana condition in order
to eliminate the doubling of the fermionic spectrum. If we had
started with $F$ complex, we should have again a chiral theory
since in this case $\overline{h}^{R}_{k}$ is different from
$h^{L}_{k}$ $(\sigma_{d}$ non self-conjugate). Starting with $F$
vector-like along with the Majorana condition will be used in our
discussions. The Majorana condition can be imposed in $D =
2,3,4+8n$ dimensions and is given by $\psi = C\overline\psi^{T}$,
where $C$ is the $D$-dimensional charge conjugation matrix.
Majorana and Weyl conditions are compatible in $D=4n+2$
dimensions. Then in our case if we start with Weyl-Majorana
spinors in $D=4n+2$ dimensions we force $f_{R}$ to be the charge
conjugate to $f_{L}$, thus arriving in a theory with fermions only
in $f_{L}$. Furthermore if $F$ is to be real, then we have to have
$D=2+8n$, while for $F$ pseudoreal $D=6+8n$.

The potential is obtained from the internal components of the
field strength. In CSDR they are given by
\begin{equation}
F_{ab} = f_{ab}^{C}\phi_{C}-[\phi_{a},\phi_{b}],
\end{equation}
so the potential is
\begin{eqnarray}
V(\phi) &=& - \frac{1}{4} g^{ac}g^{bd}Tr( f _{ab}^{C}\phi_{C} -
[\phi_{a},\phi_{b}] )\nonumber\\&&\times (f_{cd}^{D}\phi_{D} -
[\phi_{c},\phi_{d}]).
\end{eqnarray}
The minimization of the potential is in general a difficult
problem. If however $S$ has an isomorphic image $S_{G}$ in $G$
which contains $R_{G}$, the minimum is zero. Furthermore, the
four-dimensional gauge group $H$ breaks further by these non-zero
vacuum expectation values of the Higgs fields to the centralizer
$K$ of the image of $S$ in $G$, i.e. $K=C_{G}(S)$ \cite{Harnad}.
More generally it can be proven \cite{Review} that dimensional
reduction over a symmetric coset space always gives a potential of
spontaneous breaking form.

The effective action in four dimensions can be written as
\begin{eqnarray}\label{efact}
S&=&C \int d^{4}x \biggl( -\frac{1}{4} F^{t}_{\mu
\nu}{F^{t}}^{\mu\nu}\nonumber\\&&-\frac{1}{2}(D_{\mu}\phi_{\alpha})^{t}
(D^{\mu}\phi^{\alpha})^{t}
+V(\phi)\nonumber\\&&+\frac{i}{2}\overline{\psi}\Gamma^{\mu}D_{\mu}\psi-\frac{i}{2}
\overline{\psi}\Gamma^{a}D_{a}\psi\biggr),
\end{eqnarray}
where
\begin{equation}
D_{\mu} = \partial_{\mu} - A_{\mu}
\end{equation}
is the spacetime part of the gauge covariant derivative and
\begin{equation}
D_{a}=
\partial_{a}- \theta_{a}-\phi_{a},
\end{equation}
is the internal part of the gauge covariant derivative, with
$$\theta_{a}=
\frac{1}{2}\theta_{abc}\Sigma^{bc},$$ the spin connection of the
coset space. $C$ is the volume of the coset space.

Note that the second fermion term in eq.~(\ref{efact}) can be
written as
\begin{equation}\label{eq:flang}
L_{Y}=\frac{i}{2}\overline{\psi}\Gamma^{a}\nabla_{a}\psi+
\overline{\psi}V\psi ,
\end{equation}
where
\begin{eqnarray}\label{eq:nabla}
\nabla_{a}& =& - \partial_{a} +
\frac{1}{2}f_{ibc}e^{i}_{\gamma}e^{\gamma}_{a}\Sigma^{bc} +
\phi_{a}, \nonumber\\
 V&=&\frac{i}{4}\Gamma^{a}G_{abc}\Sigma^{bc}.
\end{eqnarray}
In eq.~(\ref{eq:nabla}) we have used the full connection with
torsion,
\begin{eqnarray}
\theta_{c b}^{a} &=& -
f_{ib}^{a}e^{i}_{\alpha}e^{\alpha}_{c}-(D_{cb}^{a} +
\frac{1}{2}\Sigma_{c b}^{a})\nonumber \\ &=& -
f_{ib}^{a}e^{i}_{\alpha}e^{\alpha}_{c}- G_{cb}^{a}
\end{eqnarray}
with
\begin{equation}
D_{cb}^{a} = g^{ad}\frac{1}{2}[f_{db}^{e}g_{ec} + f_{cb}^{e}g_{de}
- f_{cd}^{e}g_{be}],
\end{equation}
and $\Sigma_{c b}^{a}$ the contorsion. The general choice of the
contorsion tensor is
\begin{equation}
\Sigma_{abc}= 2\tau(D_{abc} +D_{bca} - D_{cba}),
\end{equation}
where $\tau$ is a free parameter. Furthermore the constraints
imply, that $\nabla_{a}= \phi_{a}$, when $\nabla_{a}$ acts on a
spinor field, and the term
$\frac{i}{2}\overline{\psi}\Gamma^{a}\nabla_{a}\psi $ in
eq.~(\ref{eq:flang}) is exactly the Yukawa term. The last term,
$V= \frac{i}{4}\Gamma^{a}G_{abc}\Sigma^{bc} $ gives mass to the
gaugino as will be explained in the next section.

\section{Coset Space Dimensional Reduction over symmetric coset
spaces}\label{sec:symm} Recently we have examined the CSDR of
supersymmetric gauge theories in ten dimensions over all
six-dimensional coset spaces \cite{Megalo}. Our results are that
when the  coset space is symmetric, there is no remnant of the
supersymmetric spectrum in the four-dimensional theory. Here we
present two examples representing this case, one is instructive
and the other is a promising candidate to become a realistic
model.
\subsection{CSDR over $SO(7)/SO(6)$}\label{subsec:sphere}
The ten-dimensional gauge group that we consider is $G = E_{8}$.
First we examine the dimensional reduction over $S/R =
SO(7)/SO(6)$, which is a symmetric coset space. To apply the rules
we need the embedding of $R = SO(6)$ in $G = E_{8}$ and the
decomposition of the adjoint of $G$,
\begin{eqnarray}
E_{8} &\supset& SO(6) \times SO(10)\nonumber\\ 248 &=&
(15,1)+(1,45)+(6,10)\nonumber\\&&+(4,16)+(\overline{4},\overline{16}).
\end{eqnarray}
Then the four-dimensional gauge group is
$H=C_{E_{8}}(SO(6))=SO(10)$.

To find the $R~=~SO(6)$ content of $SO(7)/SO(6)$ vector and spinor
we need the decompositions
\begin{eqnarray}
SO(7) &\supset& SO(6) \nonumber \\
21 &=& 15+6,
\end{eqnarray}
and
\begin{eqnarray}
SO(6) &\supset& SO(6) \nonumber \\ 4 &=& 4.
\end{eqnarray}
Therefore the $R=SO(6)$ content of the vector and spinor of $SO(7
)/SO(6)$ are $6$ and $4$ respectively.

The rules stated in section 2 are telling us that from the
ten-dimensional gauge field $A^{a}_{M}$ we obtain in four
dimensions the gauge field $A^{\alpha\beta}_{\mu}$ and the scalars
$\chi^{m}$, where $a$ is a $248$ $E_{8}$-index, $\alpha\beta$ a
$45$ and $m$ a $10$ $SO(10)$-index. From the ten-dimensional
gaugino  $\psi^{a}$, we obtain the four-dimensional left-handed
matter fermions $\psi^{s}$, belonging to $16$ of $SO(10)$. In
summary
\begin{eqnarray}
A^{a}_{M} &\longrightarrow& A^{\alpha\beta}_{\mu} ,
\chi^{m},\nonumber\\ \psi^{a} &\longrightarrow& \psi^{s}.
\end{eqnarray}

Note that there is no supersymmetric partner of the
four-dimensional gauge field and the scalar matter in four
dimensions transforms as a $10$-plet, while fermion matter as a
left-handed $16$-plet, therefore the spectrum of the
four-dimensional theory is non-supersymmetric.
\subsection{CSDR over $CP^{2} \times S^{2}$}
Choosing $G=E_{8}$ and $S/R=CP^{2} \times S^{2}$ which is the
coset $SU(3) \times SU(2)/SU(2) \times U(1) \times U(1)$ we have
an interesting  and promising example. The embedding of $R=SU(2)
\times U(1) \times U(1)$ in $E_{8}$ is given by the decomposition
$$E_{8} \supset SU(2) \times U(1) \times U(1) \times SO(10)$$
\begin{eqnarray}
248&=&(1,45)_{(0,0)}+(3,1)_{(0,0)}+(1,1)_{(0,0)}\nonumber\\&&+(1,1)_{(0,0)}+(1,1)_{(2,0)}
+(1,1)_{(-2,0)}\nonumber\\&&+(2,1)_{(1,2)}+(2,1)_{(-1,2)}\nonumber\\&&+(2,1)_{(-1,-2)}+(2,1)_{(1,-2)}
\nonumber\\&&+(1,10)_{(0,2)}+(1,10)_{(0,-2)}\nonumber\\&&+(2,10)_{(1,0)}+(2,10)_{(-1,0)}
\nonumber\\&&+(2,16)_{(0,1)}+(1,16)_{(1,-1)}\nonumber\\&&+(1,16)_{(-1,-1)}
+(2,\overline{16})_{(0,-1)}\nonumber\\&&+(1,\overline{16})_{(-1,1)}+(1,\overline{16})_
{(1,1)}.
\end{eqnarray}
The four-dimensional gauge group is $H=C_{E_{8}}(SU(2) \times U(1)
\times U(1))=SO(10) \times U(1) \times U(1)$. The vector and
spinor content under $R$ of the specific coset are,
$1_{0,2a}+1_{0,-2a}+2_{b,0}+2_{-b,0}$ and
$1_{b,-a}+1_{-b,-a}+2_{0,a}$ respectively. Choosing $a=b=1$ we
find that the scalar fields of the four-dimensional theory
transform as $10_{(0,2)}$, $10_{(0,-2)}$, $10_{(1,0)}$,
$10_{(-1,0)}$ under $H$. Also, we find that the fermions of the
four-dimensional theory are the following left-handed multiplets
of $H$: $16_{(-1,-1)}$, $16_{(1,-1)}$, $16_{(0,1)}$. Therefore
altogether we find
\begin{eqnarray}
A^{a}_{M} &\longrightarrow& A^{\alpha\beta}_{\mu},\ \chi_{1}^{m},\
\chi_{2}^{m},\  \chi_{3}^{m},\  \chi_{4}^{m} \nonumber\\ \psi^{a}
&\longrightarrow&  \psi_{1}^{s},\  \psi_{2}^{s},\  \psi_{3}^{s},
\end{eqnarray}
where the index conventions are as in the previous
subsection~\ref{subsec:sphere}. Worth noting are the
non-supersymmetric spectrum as in the previous example as well as
the three fermion generations which  is a clear improvement as
compared to the previous case.

Analogous conclusions concerning supersymmetry can be drawn by
examining the rest six-dimensional symmetric coset spaces
\cite{Megalo}.
\section{Soft SUSY breaking by CSDR over non-symmetric coset
spaces}\label{sec:non-symm} In the present section we examine the
reduction of a ten-dimensional $E_{8}$ gauge theory over the three
non-symmetric coset spaces. We find that a softly broken
supersymmetric four-dimensional gauge theory is obtained.
\subsection{CSDR over $G_{2}/SU(3)$}
As first example of CSDR over a non-symmetric coset space, we
present the case of $G_{2}/SU(3)$ \cite{Pman1}. The rest two
six-dimensional non-symmetric coset spaces i.e. $Sp(4)/(SU(2)
\times U(1))_{non.max}$ and $SU(3)/U(1) \times U(1)$, are examined
in the following subsections \ref{subsec:Sp} and \ref{subsec:SU}
according to refs.~\cite{Pman2,Megalo}.

We start again with $G=E_{8}$ in ten dimensions, $S/R=
G_{2}/SU(3)$, and $R=SU(3)$. Now we need the decomposition
\begin{eqnarray}
E_{8} &\supset& SU(3) \times E_{6} \nonumber \\ 248 &=& (8,1) +
(1,78) + (3,27) + (\overline{3},\overline{27}) ,
\end{eqnarray}
from which we find  that the four-dimensional gauge group is
$H=C_{SU(3)}(E_{8})=E_{6}$.\\ Next, to find the $R=SU(3)$ content
of $G_{2}/SU(3)$ vector and spinor, we examine the decompositions
\begin{eqnarray}
G_{2} &\supset& SU(3) \nonumber\\ 14&=&8+3 + \overline{3}
\end{eqnarray}
and
\begin{eqnarray}
SO(6)&\supset& SU(3)\nonumber\\ 4&=&1+3
\end{eqnarray}
from which we conclude that the $R=SU(3)$ content of $G_{2}/SU(3)$
vector and spinor are $3 + \overline{3} $ and $1+3$ respectively.
Then the CSDR rules determine that from the ten-dimensional gauge
field $A^{a}_{M}$ we obtain in four dimensions the gauge field
$A^{\alpha}_{\mu}$ and the scalars  $\beta^{i}$, where $a$ is a
$248$ $E_{8}$-index, $\alpha$ a $78$ and $i$ a $27$-$E_{6}$ index.
From the ten-dimensional gaugino  $\psi^{a}$ we obtain the
four-dimensional gaugino  $\lambda^{\alpha}$ and the left-handed
matter fermions  $\psi^{i}_{\beta}$. In summary we find
\begin{eqnarray}
A^{a}_{M} \longrightarrow A^{\alpha}_{\mu} , \beta^{i},\nonumber\\
\psi^{a} \longrightarrow \lambda^{\alpha}, \psi^{i}_{\beta}.
\end{eqnarray}

Note that the surviving fields are organized as four-dimensional
${\cal N}=1$ vector and chiral multiplets $V^{\alpha}$ and $
B^{i}$.

Moreover the scalar potential, using the metric of $G_{2}/SU(3)$,
\begin{equation}
g_{ab}=R^{2}\delta_{ab},
\end{equation}
where $R$ is the radius of $G_{2}/SU(3)$, was found \cite{Pman1}
to be
\begin{eqnarray}
V(\beta)&=& \frac{8}{R^{4}}- \frac{40}{3R^{2}}\beta^{2} -
\left[\frac{4}{R}d_{ijk}\beta^{i}\beta^{j}\beta^{k}
+h.c\right]\nonumber
\\& &+\beta^{i}\beta^{j}d_{ijk}d^{klm}\beta_{l}\beta_{m} \nonumber\\
& &+\frac{11}{4}\sum_{\alpha}\beta^{i}(G^{\alpha})_{i}^{j}
\beta_{j}\beta^{k}(G^{\alpha})_{k}^{l}\beta_{l}.
\end{eqnarray}
In turn it was found that the $F$-terms are obtained from the
superpotential
\begin{equation}
{\cal W} (B) =\frac{1}{3}d_{ijk}B^{i}B^{j}B^{k},
\end{equation}
where $d_{ijk}$ is the $E_{6}$-symmetric invariant tensor
\cite{Kephart}.

Similarly the D-terms were found to be
\begin{equation}
D^{\alpha}
=\sqrt{\frac{11}{2}}\beta^{i}(G^{\alpha})^{j}_{i}\beta_{j},
\end{equation}
where $(G^{\alpha})^{j}_{i}$ are representation  matrices for the
$27$ of $E_{6}$. The rest terms of the scalar potential that are
not obtained from $F$- or $D$- terms belong to the scalar Soft
Supersymmetry Breaking (SSB) part of the Lagrangian, given by
\begin{equation}
L_{SSB}=-\frac{40}{3R^{2}}\beta^{2}-\left[\frac{4}{R}d_{ijk}\beta^{i}\beta^{j}\beta^{k}+h.c\right].
\end{equation}
The SSB sector is completed by the gaugino Mass $M$, which is
obtained from the $V$ operator of eqs.(\ref{eq:flang}),
(\ref{eq:nabla}), and was found to be in the present case
\begin{equation}
M=(1+3\tau)\frac{6}{\sqrt{3}R}.
\end{equation}
\subsection{CSDR over $Sp(4)/(SU(2) \times
U(1))_{non.max}$}\label{subsec:Sp} This time we have again
$G=E_{8}$ in ten dimensions but $S/R=Sp(4)/(SU(2) \times
U(1))_{non.max}$. The decomposition to be used is
\begin{eqnarray}
E_{8} &\supset& SU(2) \times U(1) \times E_{6}\nonumber\\
 248 &=&
(3_{0},1)+(1_{0},1)+(1_{0},78)\nonumber\\&&+(2_{3},1)+(2_{-3},1)\nonumber\\
&&+(2_{1},27)+(2_{-1},\overline{27})\nonumber\\&&+(1_{-2},27)
+(1_{2},\overline{27}).
\end{eqnarray}
Thus the four-dimensional gauge group is
\begin{equation}
H=C_{E_{8}}(SU(2) \times U(1))= E_{6} \times U(1).
\end{equation}
Now the $R~=~SU(2) \times U(1)$ content of $Sp(4)/(SU(2) \times
U(1))_{non-max.}$ vector and spinor are
\begin{equation}
2_{1}+2_{-1}+1_{2}+1_{-2},
\end{equation}
and
\begin{equation}
2_{1}+1_{0}+1_{-2},
\end{equation}
respectively. From the CSDR rules we read that from the
higher-dimensional gauge field we obtain the four-dimensional
gauge fields, $A^{\alpha}_{\mu}$, $A_{\mu}$, with $\alpha$ a $78$
$E_{6}$-index, and two complex scalar fields $\beta^{i}$,
$\gamma^{i}$, with $i$ a $27$ $E_{6}$-index. From the
ten-dimensional gaugino we obtain the four-dimensional gaugino $
\lambda^{\alpha}$, $\lambda$, and two spinors $\psi^{i}_{\beta}$,
and  $\psi^{i}_{\gamma}$ belonging to the $27$ of $E_{6}$, i.e.
\begin{eqnarray}
A^{a}_{M} &\longrightarrow& A^{\alpha}_{\mu}, A_{\mu}, \beta^{i},
\gamma^{i},\nonumber\\ \psi^{a} &\longrightarrow&
\lambda^{\alpha}, \lambda, \psi^{i}_{\beta}, \psi^{i}_{\gamma}.
\end{eqnarray}
Note that the surviving fields can be organized as two vector
multiplets $V^{\alpha}$, $V$ and two chiral multiplets $ B^{i}$,
and $C^{i}$.

As in the previous case, the Lagrangian is ${\cal N}=1$
supersymmetric supplemented by a soft SSB part. The F-terms of the
supersymmetric part are obtained from superpotential
\begin{equation}
{\cal W}(B^{i},C^{j})= \sqrt{\frac{5}{7}}d_{ijk}B^{i}B^{j}C^{k},
\end{equation}
while the scalar SSB part is,
\begin{eqnarray}
{ \cal L}_{scalarSSB}= -\frac{6}{R_{1}^{2}}\beta^{i}\beta_{i}
-\frac{4}{R_{2}^{2}}\gamma^{i}\gamma_{i}\nonumber\\ +
\bigg[4\sqrt{\frac{10}{7}}R_{2}
\bigl(\frac{1}{R_{2}^{2}}+\frac{1}{2R_{1}^{2}}\bigr)
d_{ijk}\beta^{i}\beta^{j}\gamma^{k} + h.c \biggr].
\end{eqnarray}
$R_{1}$ and $R_{2}$ are the coset space scales coming from the
metric,
\begin{equation}
g_{ab}=diag(R_{1}^{2}, R_{1}^{2}, R_{2}^{2}, R_{2}^{2}, R_{1}^{2},
R_{1}^{2}).
\end{equation}
Finally the gaugino Mass calculated from the $V$ operator is,
\begin{equation}
M=(1+3\tau)\frac{R_{2}^{2}+2R_{1}^{2}}{8R_{1}^{2}R_{2}}.
\end{equation}
\subsection{CSDR over $SU(3)/U(1) \times U(1)$}\label{subsec:SU}
The $G=E_{8}$, ten-dimensional theory is now reduced over the last
non-symmetric coset space $S/R=SU(3)/U(1) \times U(1)$. From the
decomposition
\begin{eqnarray}
E_{8} &\supset& U(1)_{1} \times U(1)_{2} \times E_{6}\nonumber\\
 248 &=&1_{(0,0)}+1_{(0,0)}+1_{(3,\frac{1}{2})}\nonumber\\&&+1_{(-3,\frac{1}{2})}+
1_{(0,-1)}+1_{(0,1)}\nonumber\\&&+1_{(-3,-\frac{1}{2})}+1_{(-3,-\frac{1}{2})}\nonumber\\&&+
78_{(0,0)}+27_{(3,\frac{1}{2})}\nonumber\\&&+27_{(-3,\frac{1}{2})}+27_{(0,-1)}\nonumber\\
&&+\overline{27}_{(-3,-\frac{1}{2})}+\overline{27}_{(3,-\frac{1}{2})}\nonumber\\&&
+\overline{27}_{(0,1)},
\end{eqnarray}
we conclude that the four-dimensional gauge group is,
\begin{equation}
 H=C_{E_{8}}(U(1)_{1} \times U(1)_{2}) = U(1)_{1} \times U(1)_{2}
\times E_{6}.
\end{equation}
The $R=U(1) \times U(1)$ content of $SU(3)/U(1) \times U(1)$
vector and spinor are
\begin{eqnarray}
(3,\frac{1}{2})+(-3,\frac{1}{2})
+(0,-1)+(-3,-\frac{1}{2})\nonumber\\+(3,-\frac{1}{2})+(0,1),
\end{eqnarray}
and
\begin{equation}
(0,0)+(3,\frac{1}{2})+(-3,\frac{1}{2}) +(0,-1),
\end{equation}
respectively.

Applying the CSDR rules we find that from the ten-dimensional
gauge fields we obtain the four-dimensional gauge fields
$A^{\alpha}_{\mu}$, with $\alpha$ a $78$ $E_{6}$-index,
$A_{(1)\mu}$, $A_{(2)\mu}$, the two $U(1)$ gauge fields, three
scalars $\alpha^{i}$, $\beta^{i}$, $\gamma^{i}$, with $i$ a $27$
index, and three $E_{6}$ singlets but charged under the $U(1)$s,
$\alpha$, $\beta$, and $\gamma$. From the ten-dimensional gaugino
we obtain the supersymmetric partners of the above scalars, i.e.
\begin{eqnarray}
A^{a}_{M} &\longrightarrow& A^{\alpha}_{\mu}, A_{(1)\mu},
A_{(2)\mu},\nonumber\\ &&\alpha^{i}, \beta^{i}, \gamma^{i},
\alpha, \beta, \gamma,\nonumber
\\ \psi^{a} &\longrightarrow& \lambda^{\alpha}, \lambda_{(1)},
\lambda_{(1)},\nonumber\\ &&\psi^{i}_{\alpha}, \psi^{i}_{\beta},
\psi^{i}_{\gamma},\psi_{\alpha}, \psi_{\beta}, \psi_{\gamma}.
\end{eqnarray}
Therefore the surviving fields are organized as three vector
multiplets, $V^{\alpha}$, $V_{(1)}$, $V_{(2)}$, and six chiral
multiplets, $A^{i}$, $B^{i}$, $C^{i}$, $A$, $B$, $C$.

As in the previous two cases the Lagrangian is softly broken
supersymmetric. The $F$-terms come from the superpotential
\begin{eqnarray}
{ \cal W}(A^{i},B^{j},C^{k},A,B,C)
&=&\sqrt{40}d_{ijk}A^{i}B^{j}C^{k}\nonumber\\ &&+ \sqrt{40}ABC.
\end{eqnarray}
The scalar SSB part of the Lagrangian is,
\begin{eqnarray}\label{eq:SSB}
\lefteqn{{\cal L}_{scalarSSB}=\biggl(
\frac{4R_{1}^{2}}{R_{2}^{2}R_{3}^{2}}-\frac{8}{R_{1}^{2}}
\biggr)\alpha^{i}\alpha_{i}}\nonumber\\ && +\biggl(
\frac{4R_{1}^{2}}{R_{2}^{2}R_{3}^{2}}-\frac{8}{R_{1}^{2}}
\biggr)\overline{\alpha}\alpha \nonumber\\ & &
+\biggl(\frac{4R_{2}^{2}}{R_{1}^{2}R_{3}^{2}}-\frac{8}{R_{2}^{2}}\biggr)
\beta^{i}\beta_{i}
+\biggl(\frac{4R_{2}^{2}}{R_{1}^{2}R_{3}^{2}}-\frac{8}{R_{2}^{2}}\biggr)
\overline{\beta}\beta \nonumber\\ & &
+\biggl(\frac{4R_{3}^{2}}{R_{1}^{2}R_{2}^{2}}
-\frac{8}{R_{3}^{2}}\biggr)\gamma^{i}\gamma_{i}
+\biggl(\frac{4R_{3}^{2}}{R_{1}^{2}R_{2}^{2}}
-\frac{8}{R_{3}^{2}}\biggr)\overline{\gamma}\gamma \nonumber\\ & &
+\biggl[\sqrt{2}80\biggl(\frac{R_{1}}{R_{2}R_{3}}+\frac{R_{2}}{R_{1}
R_{3}}+\frac{R_{3}}{R_{2}R_{1}}\biggr)\nonumber\\&& \times \
d_{ijk}\alpha^{i}\beta^{j}\gamma^{k} \nonumber \\ &
&+\sqrt{2}80\biggl(\frac{R_{1}}{R_{2}R_{3}}+\frac{R_{2}}{R_{1}
R_{3}}+\frac{R_{3}}{R_{2}R_{1}}\biggr)\nonumber\\&& \times \
\alpha\beta\gamma+ h.c\biggr].
\end{eqnarray}
The coset space radii, $R_{1}$, $R_{2}$, $R_{3}$ are entering the
above formula (\ref{eq:SSB}) through the metric, which takes the
following form for the present coset $SU(3)/U(1) \times U(1)$
\begin{equation}
g_{ab}=diag(R^{2}_{1}, R^{2}_{1}, R^{2}_{2}, R^{2}_{2}, R^{2}_{3},
R^{2}_{3}).
\end{equation}

The SSB sector is completed by the gaugino mass,
\begin{equation}
M=(1+3\tau)\frac{(R_{1}^{2}+R_{2}^{2}+R_{3}^{2})}{8\sqrt{R_{1}^{2}R_{2}^{2}R_{3}^{2}}}.\nonumber
\end{equation}

Thus we have established that besides the supersymmetric spectrum,
we obtain a softly broken supersymmetric Lagrangian by CSDR over a
non-symmetric coset space.
\section{Conclusions}\label{sec:concl}
The CSDR was originally introduced as a scheme which, making use
 of higher dimensions, incorporates in a
unified manner the gauge and the ad-hoc Higgs sector of the
spontaneously broken gauge theories in four dimensions
\cite{Witten,Manton}. Next fermions were introduced in the scheme
and the ad-hoc Yukawa interactions have also been included in the
unified description \cite{Slansky,Chapline}.

Of particular interest for the construction of realistic theories
in the framework of CSDR are the following virtues that
complemented the original suggestion: (i) The possibility to
obtain chiral fermions in four dimensions resulting from
vector-like representations of the higher-dimensional gauge theory
\cite{Chapline,Review}. This possibility can be realized due the
presence of non-trivial background gauge configurations which are
introduced by the CSDR constructions \cite{Salam}, (ii) The
possibility to deform the metric of certain non-symmetric coset
spaces and thereby obtain more than one scales
\cite{Review,Hanlon}, (iii) The possibility to use coset spaces,
which are multiply connected. This can be achieved by exploiting
the discrete symmetries of the S/R \cite{Kozimirov,Review}. Then
one might introduce topologically non-trivial gauge field
\cite{Zoupanos} configurations with vanishing field strength and
induce additional breaking of the gauge symmetry. It is the
Hosotani mechanism \cite{Hosotani} applied in the CSDR.

In the above list recently has been added the interesting
possibility that the popular softly broken supersymmetric
four-dimensional chiral gauge theories might have their origin in
a higher-dimensional supersymmetric theory with only vector
supermultiplet \cite{Pman1,Pman2,Megalo}, which is dimensionally
reduced over non-symmetric coset spaces.

Let us also note that the current discussion on the
higher-dimensional theories with large extra dimensions
\cite{Anton}, provides a new framework to examine further the CSDR
since the classical treatment used in CSDR is justified in the
case of large radii which are far away from the scales that the
quantum effects of gravity are important.

On the other hand we should also note that the effective field
theories resulting from compactification of higher-dimensional
theories contain also towers of massive higher harmonic
(Kaluza-Klein) excitations, whose contributions at the quantum
level alter the behaviour of the running couplings from
logarithmic to power \cite{Taylor}. As a result the traditional
picture of unification may change drastically \cite{Dienes,Kubo}.
Combining the quantum behaviour of the higher-dimensional theories
\cite{Dienes,Kubo}, with the unification at the classical level of
the gauge-Higgs and the gauge-Yukawa sectors that can be achieved
in the CSDR scheme, one might hope to achieve a {\it reduction of
couplings} of the SM by reducing the dimensions of a
higher-dimensional theory.

\end{document}